\documentstyle[]{article}

\begin{document}
                     
\noindent
\centerline{{\large \bf LIFE EXTINCTIONS BY}} 
\centerline{{\large \bf COSMIC RAY JETS}} 

\medskip
\noindent
\centerline{{\bf Arnon Dar, Ari 
Laor, and Nir J. Shaviv}}

\noindent
\centerline {Department of Physics and Space Research Institute}
\centerline {Israel 
Institute of Technology, Haifa 32000, Israel.}

\begin{abstract}  
\noindent
High energy cosmic ray jets from nearby mergers or accretion induced
collapse (AIC) of neutron stars (NS) that hit the atmosphere can produce
lethal fluxes of atmospheric muons at ground level, underground and
underwater, destroy the ozone layer and radioactivate the environment.
They could have caused most of the massive life extinctions on planet
Earth in the past $600~My$.  Biological mutations due to ionizing
radiations could have caused the fast appearance of new species after the
massive extinctions. An early warning of future extinctions due to NS
mergers may be obtained by identifying, mapping and timing all the nearby
binary neutron stars systems. A warning of an approaching cosmic ray burst
from a nearby NS merger/AIC may be provided by a very intense gamma ray
burst which preceeds it. 

\end{abstract} 

\section{Introduction}

The early history of single-celled organisms during the Precambrian, 4560
to 570 million years  ($My$) ago, is poorly known. Since the end of the
Precambrian the diversity of both marine and continental life increased
exponentially. Analysis of fossil record of microbes, algae, fungi,
protists, plants and animals shows that this diversification was
interrupted by five major mass extinction ``events'' and some smaller
extinction peaks [1] . The ``big five'' mass extinctions occurred in the
Late Ordovician, Late Devonian Late Permian, Late Triassic and
end-Cretaceous and included both marine and continental life. The largest
extinction occurred about $251~My$ ago at the end of the Permian
period. The global species extinction ranged then between 80\% to 95\%,
much more than, for instance, the end-Ordovician extinction $439~My$ ago 
which eliminated 57\% of marine genera, or the
Cretaceous-Tertiary extinction $64~My$ ago which killed the
dinosaurs and claimed 47\% of existing genera [2]. In spite of intensive
studies it is still not known what caused the mass extinctions, how quick
were they and whether they were subject to regional variations.  Many
extinction mechanisms have been proposed but no single mechanism seems to
provide a satisfactory explanation of both the marine and continental
extinction levels, the biological extinction patterns and the repetition
rate of the mass extinctions [1,2]. These include astrophysical extinction
mechanisms, such as meteoritic impact that explains the iridiun
anomaly which was found at the Cretaceous/Tertiary boundary [3] but has
not been found in all the other extinctions [4], supernova explosions [5] and
gamma ray bursts [6], which do not occur close enough at sufficiently high
rate to explain the observed rate of mass extinctions.

In this paper we propose that high energy cosmic ray jets (CRJs) from
mergers or accretion induced collapse (AIC) of neutron stars (NS) that hit
the atmosphere of an Earth-like planet can produce lethal fluxes of
atmospheric muons at ground level, underground and underwater, destroy the
ozone layer and radioactivate the environment. Nearby NS mergers/AIC can
explain the massive extinction on the ground, underground and underwater
and the higher survival levels of radiation resistant species and terrain
sheltered species in the five ``great'' extinctions in the past $600~My$. 
More distant galactic mergers/AIC can cause smaller extinctions.
Biological mutations due to ionizing radiations may explain the fast
appearance of new species after massive extinctions.  Intense cosmic ray
bursts enrich rock layers with detectable traces of cosmogenically
produced radioactive nucleides such as $^{129}$I, $^{146}$Sm, $^{205}$Pb
and $^{244}$Pu. Tracks of high energy particles in rock layers on Earth
and on the moon may also contain records of intense cosmic irradiations.
An early warning of future extinctions due to neutron star mergers can be
obtained by identifying, mapping and timing all the nearby binary neutron
stars systems. A final warning of an approaching CRJ from a nearby neutron
stars merger is provided by a very intense gamma ray burst a few days
before the arrival of the CRJ.

\section{Cosmic Ray Jets From NS Mergers}

Three NS-NS binaries are presently known within the galactic disc;  
B1913+16 [7] at a distance of $D\sim 7.3~ kpc$, B203+46 [8] at $D\sim
2.3~kpc$, and B1534+12 [9] at $D\sim 0.5~kpc$, and  B2127+11C [10] in
the globular cluster M15 at $D\sim 10.6~kpc$. Although unseen as radio
pulsars, the companion stars in these systems have been identified as
neutron stars from their mass which is inferred via measurements of the
relativistic periastron advance using standard pulse timing techniques
[11]. The continuous energy loss through gravitational radiation brings
the two NS closer and closer until they merge. It is believed that their
final merger releases an enormous amount of gravitational binding energy,
$\sim M_\odot c^2$, in a few ms, in the form of gravitational waves,
neutrinos and kinetic energy of relativistic ejecta.  It was suggested
that the neutrinos and/or the relativistic ejecta from NS merger/AIC in
distant galaxies produce the mysterious gamma ray bursts (GRBs), that
occur at a rate of about one per day, although the mechanism which
converts their energy into gamma rays is still not clear [12]. 

It is also believed that due to strong gravitational tidal forces the
final NS-NS or NS-BH merger proceeds through the formation of an accretion
disc. Observations seem to indicate that highly collimated jets are
ejected by all systems where matter is undergoing disc accretion onto a
compact central object [13,14]. They also indicate that the jet kinetic
energy is a considerable fraction of the accretion power and that the jets
reach enormous distances without being deflected by galactic or
intergalactic magnetic fields: The highly relativistic and highly
collimated jets from active galactic nuclei (AGN), which are believed to
be powered by mass accretion onto a massive black hole at a typical rate
of $(\sim M_\odot~ y^{-1})$, reach a distance up to a million light years
before disruption [13].  Highly collimated relativistic matter that is
ejected sporadicly by microquasars (superluminal galactic sources such as
GRS 1915+105 [14, 15], GRO J165-40 [16]) and by X-ray sources such as SS
433 [17] and Cygnus X-3 [18], which are are close binary systems where
mass is accreted onto a neutron star or a stellar black hole from a
companion star, seem to reach hundreds of light years before disruption. 
The exact mechanism by which the gravitational and electromagnetic fields
around these accreting and rotating compact objects produce the highly
relativistic jets is still unknown. However, the instantaneous accretion
rates ($\gg M_\odot s^{-1}$) and the strength of the magnetic fields which
are involved in the final stage of merger/AIC of compact stellar objects
in compact binary systems are probably many orders of magnitude larger
than those encountered in AGN. Therefore, it is natural to expect that
highly relativistic jets are also ejected in mergers/AIC of compact
stellar objects, are highly collimated and reach hundreds of light years,
or more, before disruption.  These jets may produce the cosmological gamma
ray bursts by internal radiation and/or interaction with the external
medium. After disruption they are isotropized by the galactic magnetic
field and form the galactic cosmic rays. If they hit an Earth-like planet
before disruption, they can devastate all forms of life on it. 

\section{Constraints From GRBs and Cosmic Rays}

Jets ejected by NS mergers/AIC in distant galaxies may explain
cosmological GRBs [19]. In view of the uncertainties in modeling jet
ejection in NS merger/AIC, rather than relying on numerical
simulations we inferred [19] from the observed properties of
GRBs that the ejected jets
have typical Lorentz factors of $\Gamma\sim 10^3$, beaming angles
$\Delta\Omega\leq 1/100$ similar to those observed/estimated for AGN and
microquasars, and ejected mass $\Delta M\sim (dM/d\Omega)\Delta \Omega\leq
10^{-4} M_\odot$ (i.e., released kinetic energy bounded by $E_K=\Gamma
\Delta Mc^2 < 0.1M_\odot c^2\sim 2\times 10^{53} erg $). 

\noindent 
The finite life times of close binaries due to gravitational radiation
emission have been estimated [12] from the observed binary period, orbital
eccentricity and the masses of the pulsar and its companion. They have
been used to estimate that the NS-NS merger rate in the Milky Way (MW)
disc is [20] $R_{MW}\sim 10^{-4}-10^{-5}~y^{-1}$.  Beamed ejection from NS
mergers/AIC requires merger/AIC rate of compact objects in the entire
Universe (at GRBs redshifts) of the order of $10^5-10^6~y^{-1}$. It is
larger by $1/\Delta\Omega\sim 10^2$ than that required by spherical
explosions. It is, however, in good agreement with the updated
observational and theoretical estimates of the merger rates of compact
objects in the Milky Way, which yield values in the range $10^5-10^6
y^{-1}$ mergers per Universe [21], instead of initial estimates [20] of
$\sim 10^3-10^4~y^{-1}$.  Thus, the updated estimates of the NS-NS merger
rate in the Universe are consistent with the observed rate [12] of GRBs
(approximately one per day) and with the injection rate of cosmic rays in
the MW that is required in order to maintain a constant cosmic ray flux in
the MW: 

\noindent
The escape rate of cosmic rays from the MW requires an average injection
rate of $ Q_{CR}\sim 10^{41}~erg~s^{-1}$, in high energy cosmic
rays above GeV, in order to maintain a constant energy density of cosmic
rays in the MW [22]. 
This injection rate can be supplied by NS-NS mergers if the injected
jet energy per merger is $E_K \sim 10^{53}erg$ and if the 
NS-NS merger rate in the disc of the MW is  
$R_{MW}\sim 10^{-4}-10^{-5}~y^{-1}$:  
\begin{equation}
Q_{CR}^{NS} \sim R_{MW} E_K\sim 3\times10^{40}-3\times 10^{41}~erg~s^{-1}. 
\end{equation}

\section{Attenuation of CRJs}

The highly relativistic jets from quasars and microquasars 
do not seem to be attenuated efficiently in the interstellar or the 
intergalactic space. 
This is unlike the non relativistic ejecta
in supernova (SN) explosions, which is attenuated by Coulomb collisions in
the interstellar medium (ISM) over a distance of a few $pc$:  Moderately
energetic charged particles, other than electrons, lose energy in neutral
interstellar gas primarily by ionization. The mean rate of energy loss (or
stopping power) is given by the Bethe-Bloch formula
\begin{equation}
-dE/dx\approx (4\pi Z^2 n_e \beta^{-2}e^4/m_e
c^2)[(1/2)ln(2m_ec^2\beta^2\Gamma^2T_{max}/I^2)-\beta^2], 
\end{equation}
where $Ze$ is the charge of the energetic particle of mass $M_i$, velocity
$\beta c$ and total energy $E=\Gamma M_ic^2$, $n_e$ is the number of
electrons per unit volume in the medium in atoms with ionization potential
$I$, and $T_{max}=2m_ec^2\beta^2\Gamma^2 /(1+2\gamma m_e/M_i)$.  If the
interstellar gas around the SN is ionized (by the initial UV flash), then
$I$ has to be replaced by $e^2/R_D$ where $R_D=(kT/4\pi
e^2n_e(Z+1))^{1/2}$ is the Debye screening length.  Thus, for SN ejecta
with $Z\sim 1$ and $v\sim 10000~km~s^{-1}$ in an ionized interstellar
medium with a typical density of $n_{H}\sim 1~cm^{-3}$, and $kT\sim 1~eV$
the stopping distance of the ejecta due to Coulomb interactions is $x\sim
E/2(dE/dx) \sim 6~pc$. In fact, the stopping of the ejecta by Coulomb
interactions is consistent with observations of SN remnants, like SN 1006
[23], while the assumption that the ionized interstellar medium is glued
to the swept up magnetic field seems to be contradicted by some recent
observations [24]. The range increases with energy like $\beta^4$ until
nuclear collisions become the dominant loss mechanism. The range of nuclei
with $\Gamma\sim 1000$ in a typical interstellar density of $n_{H}\sim
1~cm^{-3}$, is approximately $\sim 10^{25} cm $, i.e., much larger than
galactic distances. 

Although the galactic magnetic field, $H\sim 3-5~\times 10^{-6}~ Gauss$,
results in a Larmor radius of $r_{L}=\beta\Gamma m c/qH\sim 10^{15}~cm$
for protons with $\Gamma=1000$, it does not deflect and disperse
significantly jets from NS-NS mergers at distances smaller than $\sim
1~kpc$ from the explosion. That can be concluded from the fact that
accretion jets from forming stars, microquasars and AGN reach distances of
tens, hundreds, and million light years, respectively, without significant
deflection or attenuation. Probably, because of their high particle and
energy densities the jets produce internal magnetic fields which shield
them from the interstellar magnetic field and allow them to follow almost
free balistic trajectories in the interstellar medium. 

\section{Mass Extinctions By CRJs} 

We assume that the ambient interstellar gas is not swept up with the jet.
If it were, then the jet would not reach even $\sim 10~pc$. In SN
explosions collective modes are invoked as the source of the coupling of
the ejecta to the interstellar medium, required in order to attenuate the
SN debris. As mentioned above, binary Coulomb interactions are sufficient
to produce the observed coupling in SN explosions, and coupling through
collective modes is not necessarily present. Due to internal magnetic
fields the jets are highly collimated, not deflected and probably reach
distances of $D\sim 1~kpc$, where the internal energy density
becomes compareble to the external (magnetic and radiation) energy density. 

Unattenuated jets from NS-NS mergers can be devastating to life on 
nearby planets: At a distance of $1~kpc$ their  duration is 
\begin{equation} 
\delta t\sim D/2c\Gamma^2\sim 1~day - 2~months 
\end{equation} 
for typical values of $\Gamma$ between 1000 and 100, respectively. The
time integrated energy flux of the jet at $D\sim 1~kpc$ is, typically,
$\sim 10^{12}~TeV~cm^{-2}$. Thus, the energy deposition in the
atmosphere by the jet is equivalent to the total energy deposition of
galactic cosmic rays in the atmosphere over $\sim 10^7~ y$.  However, the
typical energy of the cosmic rays in the CRJ is $\sim 1~TeV$ per nucleon,
compared with $\sim 1~GeV$ per nucleon for ordinary cosmic ray nuclei. 
Collisions of such  particles in the atmosphere generate atmospheric
cascades where a significant fraction of the CRJ energy is converted into
``atmospheric muons'' through leptonic decay modes of the produced mesons. 
Most of these muons do not decay in the atmosphere because of their high
energy, unlike most of the atmospheric muons which are produced by
ordinary cosmic rays. The average number of high energy muons produced by
nucleons of primary energy $E_p$, which do not decay in the atmosphere and
reach sea level with energy $>E_\mu$ at zenith angle $\theta<\pi/2$, is
given approximately by [25]:
\begin{equation}
<N_\mu>\sim 
(0.0145E_p[TeV])(E_p/E_\mu)^{0.757}(1-E_\mu/E_p)^{5.25}/cos\theta~.
\end{equation}
Thus a jet with energy of about $1~TeV$ per nucleon at a distance of
$1~kpc$ produces at sea level a flux of atmospheric muons of 
\begin{equation}
 I_\mu(>3~GeV)\sim 10^{12}~ cm^{-2}.          
\end{equation} 
Such muons deposit energy in matter via ionization. Their energy
deposition rate is [26] $-dE/dx\geq 2~MeV~g^{-1}cm^{-1}$.  The whole-body
lethal dose from penetrating ionizing radiation resulting in 50\%
mortality of human beings in 30 days [26] is $\leq 300~rad \sim 2\times
10^{10}/(dE/dx) \sim 10^{10}~cm^{-2}$ where $dE/dx$ is in rate is in
$MeV~g^{-1}cm^{-1}$ units.  The lethal dosages for other vertebrates can
be a few times larger while for insects they can be as much as a factor 20
larger.  Hence, a CRJ at $D\sim 1~kpc$ which is not significantly
dispersed by the galactic magnetic field produces a highly lethal burst of
atmospheric muons. Because of muon penetration, the large muon flux is
lethal for most species even deep (hundreds of meters) underwater and
underground, if the cosmic rays arrive from well above the horizon. Thus,
unlike the other suggested extraterrestrial extinction mechanisms, a CRJ
which produces a lethal burst of atmospheric muons can explain also the
massive extinction deep underwater and why extinction is higher in
shallow waters. 

\noindent Although half of the planet is in the shade of the CRJ, planet
rotation exposes a larger fraction of the planet surface to the CRJ.
Additional effects increase the lethality of the CRJ over the whole planet.
They include: 

(a) The pollution of the environment by radioactive
nuclei, produced by spallation of atmospheric and surface nuclei by shower
particles. Using the analytical methods of [27], we estimate that for
an Earth-like atmosphere, the flux of energetic nucleons which reaches the
surface is also considerable, 
\begin{equation}
I_p(>100~MeV)\sim I_n (>100~MeV)\sim 10^{10}~cm^{-2}. 
\end{equation}
Global winds spread radioactive gases in a relatively short time over 
the whole planet. 

(b) Depletion of stratospheric ozone by the reaction of ozone with nitric
oxide, generated by the cosmic ray produced electrons in the atmosphere
(massive destruction of stratospheric ozone has been observed during large
solar flares which produced energetic protons [28]). 

(c) Extensive damage to the food chain by radioactive pollution and
massive extinction of vegetation and living organisms by ionizing
radiations (the lethal radiation dosages for trees and plants are slightly
higher than those for animals but still less than the flux given by
eq. 5 for all except the most resilient species). 

\section{Signatures of CRJ Extinction}

{\bf The biological extinction pattern:} The biological extinction pattern
due to a CRJ depends on the exposure and the vulnerability of the different
species to the primary and secondary effects of the CRJ. The exposure of
the living organisms to the muon burst depends on the intensity and 
duration of
the CRJ, on its direction relative to the rotation axis of Earth (Earth
shadowing), on the local sheltering provided by terrain (canyons, mountain
shades) and by underwater and underground habitats, and on the risk
sensing/assessment and mobility of the various species. The lethality of
the CRJ depends as well on the vulnerability of the various living 
species
and vegetation to the primary ionizing radiation, to the drastic changes
in the environment (e.g., radioactive pollution and destruction of the
ozone layer) and to the massive damage and radioactive poisoning of the
food chain. Although the exact biological signature may be quite
complicated, and somewhat obscured in fossil records (due to poor or
limited sampling, deterioration of the rocks with time and dating and
interpretation uncertainties because of bioturbational smearing) it may
show the general pattern expected from a CRJ extinction. Indeed, a first
examination of the fossil records suggest that there is a clear
correlation between the extinction pattern of different species, their
vulnerability to ionizing radiation and the sheltering provided by their
habitats and the environment they live in.  For instance, insects which
are less vulnerable to radiation, were extinct only in the greatest
extinction - the end-Permian extinction $251~My$ ago.  Even then only 8
out of 27 orders were extinct compared with a global species extinction
that ranged between 80\% to 95\% [4]. Also plants which are less
vulnerable to ionizing radiation suffered lower level of extinction.
Terrain, underground and underwater sheltering against a complete
extinction on land and in deep waters may explain why certain families on
land and in deep waters were not extinct even in the great extinctions,
while most of the species in shallow waters and on the surface were
extinct [4]. Mountain shadowing, canyons, caves, underground habitats,
deep underwater habitats and high mobility may also explain why many
species like crocodiles, turtles, frogs, (and most freshwater
vertebrates), snakes, deep sea organisms and birds were little affected in
the Cretaceous/Tertiary (K/T) boundary extinction which claimed the life
of the big dinosaurs and pterosaurs. In particular, fresh underground
waters in rivers and lakes are less polluted with radioisotopes and
poisons produced by the CRJ than sea waters and may explain the survival
of freshwater amphibians. 

\noindent 
{\bf Geological Signatures:} The terrestrial deposition of the primary CRJ
nuclides or the production of stable nuclides in the atmosphere or in the
surface by the CRJ is too small to be detectable. In particular, the
proposed mechanism cannot explain the surface enrichment at the K/T
boundary by about $3\times 10^5~tons$ of iridium. Alvarez et al. [3]
suggested that the impact of extraterrestrial asteroid, with an Ir
abundance similar to that observed in early solar system Chondritic
meteorites whose Ir abundance is larger than crustal Ir abundance by $\sim
10^4$, could cause the Ir anomaly and explain the mass extinction at the
K/T boundary. However, no significant Ir enrichment was found in all other
mass extinctions. Moreover, other isotopic anomalies due to meteoritic
origin have not been found around the K/T boundary; in particular the
As/Ir and Sb/Ir ratios are three orders of magnitude greater than
chondritic values but are in accord with a mantle origin [29].  Extensive
iridium measurements showed that the anomaly does not appear as a single
spike in the record, indicative of an instantaneous event, but rather
occur over a measurable time interval of 10 to 100 ky or possibly longer
[30]. That and the high abundance of Ir in eruptive magma led to the
suggestion that the iridium anomaly is due to a global volcanic activity
over 10 to 100 kyr at the K/T boundary [30,31] which also caused the K/T
extinction: Eruptions have a variety of short term effects, including
cooling from both dust and sulfates ejected into the stratosphere, acid
rain, wildfires, release of poisonous elements and increase in ultraviolet
radiation from ozone-layer depletion. But, examination of major volcanic
eruptions in the past $100~ My$ have shown that none of them greatly
affected the diversity of regional and global life on land or in the
oceans [4]. 

A CRJ will enhance the abundance of stable cosmogenic isotopes in the
geological layer corresponding to the CRJ event, but, the enrichment may
be negligible compared to their accumulation through long terrestrial
exposure of the geological layers to galactic cosmic rays prior to the
CRJ. However, CRJ enrichment of sediments with unstable radioisotopes of
mean lifetimes much shorter than the age of the solar system, $\tau\ll
t_\odot\approx 4570 My$, but comparable to the extinction times, may be
detectable through low traces mass spectrometry. In particular, fission of
long lived terrestrial nuclei, such as $^{238}$U and $^{232}$Th, by shower
particles, and capture of shower particles by such nuclei, may lead to
terrestrial production of, e.g., $^{129}$I with $\tau=15~My$, $^{146}$Sm
with $\tau= 146~My$, $^{205}$Pb with $\tau=43~My$ and $^{244}$Pu with
$\tau=118~My$, respectively. These radioisotopes may have been buried in
underwater sediments and underground rocks which were protected from
further exposure to cosmic rays. The main background to such a CRJ
signature is the continuous deposition by cosmic rays and by meteoritic
impacts on land and sea. Cosmic rays may include these trace radioisotopes
due to nearby sources (e.g., supernova explosions) and because of
spallation of stable cosmic ray nuclei in collisions with interstellar
gas.  Meteorites may include these trace elements due to a long exposure
in space to cosmic rays. 

Finally, large enhancement of TeV cosmic ray tracks in magma from volcanic
eruptions coincident with extinctions may also provide fingerprints for
CRJ extinctions. 

\section{Rate of Mass Extinctions}

Assuming that the spatial distribution of NS binaries and NS mergers
in the MW follow the distribution of single pulsars [],
\begin{equation} 
dN\propto e^{-R^2/2R_0^2}e^{-\vert z\vert /h}RdRdz,
\end{equation} 
with a disc scale length, $R_0\sim 4.8~kpc$, and a scale height, $h> 
0.5~kpc$
perpendicular to the disc and independent of disc position, we find that
the average rate of CRJs from NS-NS mergers that reach planet Earth from
distances  $\leq 1~kpc$, 
is $\sim 10^{-8}~y^{-1}$. It is consistent with the 5 big  
extinctions which have occurred during the last $600~My$   
in the Paleozoic and Mesozoic eras. The relative strengths of these
extinctions may reflect mainly different distances from the CRJs.
Beyond $\sim 1~ kpc$ from the explosion
the galactic magnetic field begins to disperse the CRJ and suppresses
its lethality.  Such CRJs, if not too far,  can still cause 
partial extinctions at a higher rate and induce biological mutations 
which may lead to the appearance of new species.

The galactic rate of SN explosions is $\sim 100~y^{-1}.$ The range of
debris from SN explosions in the interstellar medium is shorter than
$10~pc$. The rate of SN explosions within a distance of $10~pc$ from Earth
which follows from eq. 7 is $R_{SN}(<10~pc)\approx 10^{-10}y^{-1}$. High
energy cosmic rays which, perhaps, are produced in the SN remnant by shock
acceleration, carry only a small fraction of the total explosion energy
and arrive spread in time due to their diffusive propagation in the
interstellar magnetic field.  Also neutrino and light emissions similar to
those observed in SN1987A, at a distance of a few $pc$ cannot cause a mass
extinction. 
   
\section{Conclusions} Cosmic Ray Bursts from neutron star mergers may have
caused the massive continental and marine life extinctions which
interrupted the diversification of life on our planet.  Their rate is
consistent with the observed rate of mass extinctions in the past
$570~My.$ They may be able to explain the complicated biological and
geographical extinction patterns. Biological mutations induced by the
ionizing radiations which are produced by the CRJs may explain the
appearance of completely new species after extinctions. A first
examination suggests a significant correlation between the biological
extinction pattern of different species and their exposure and
vulnerability to the ionizing radiation produced by a CRJ. The iridium
enrichment around the Cretaceous/Tertiary extinction that claimed the life
of the dinosaurs and pterosaurs cannot be due to a CRJ. It may have been
caused by intense volcanic eruptions around that extinction. Isotopic
anomaly signatures of CRJ extinctions may be present in the geological
layers which recorded the extinctions.  Elaborate investigations of the
effects of CRJs from relatively nearby neutron star mergers and their
biological, radiological and geological fingerprints are needed before
reaching a firm conclusion whether the massive extinctions during the long
history of planet Earth were caused by CRJs from neutron stars mergers. If
nearby neutron star mergers are responsible for mass extinctions, then an
early warning of future extinctions due to neutron star mergers can be
obtained by identifying, mapping and timing all the nearby binary neutron
stars systems. A final warning for an approaching CRJ from a nearby
neutron-stars merger will be provided few days before its arrival by a
gamma ray burst produced by the approaching CRJ. 

\noindent
{\bf Acknowledgement}: This research was supported in part by the 
Technion  fund for promotion of research. 

\parindent 0cm
\centerline{References}

1. M.J. Benton, Science {\bf 278}, 52 (1995) and references 
therein. 

2. D.H. Erwin, Scientific American, {\bf 275}, 72 (1996) and 
references therein.

3. L.W. Alvarez et al., Science {\bf 208}, 1095 (1980).

4. D.H. Erwin, Nature, {\bf 367}, 231 (1994) and references therein.

5. M.A. Ruderman, Science, {\bf 184}, 1079 (1974); J. Ellis, et al.,
ApJ. {\bf 470}, 1227 (1996).  

6. S.E. Thorsett, ApJ. {\bf 444}, L53 (1995).

7. R.A. Hulse \& J. Taylor, ApJ. {\bf 195}, L51 (1975).

8. G.H. Stokes et al., ApJ. {\bf 294}, L21 (1991).

9. A. Wolszcan, Nature {\bf 350}, 688 (1991). 

10. S.B. Anderson et al.,  Nature {\bf 346}, 42 (1990).

11. R.N. Manchester \& J.H. Taylor, {\it Pulsars}, (Freeman, San Francisco
1977); J.H. Taylor \& J.M. Weisberg, ApJ. {\bf 345}, 434 
(1989).

12. G.J. Fishman \& C.A.A. Meegan, Ann. Rev. Astr. 
 Ap. {\bf 33}, 415 (1995).  

13. See, e.g., P. Padovani, these proceedings and references therein.

14. See, e.g., I.F. Mirabel, these proceedings and references therein.  

15. I.F. Mirabel and L.F. Rodriguez, Nature, {\bf 371}, 46 (1994).

16. S.J. Tingay et al., Nature, {\bf 374}, 141 (1995).

17. See, e.g., B. Margon, Ann. Rev. Ast. Ap., {\bf 22}, 57, (1988)

18. R.G. Strom et al., Nature, {\bf 337}, 234 (1989).

19. N.J. Shaviv, \& A. Dar,  Submitted to PRL, Astro-ph 9606032. 

20. E.S. Phinney, ApJ. {\bf 380}, L17 (1991);  
R. Narayan, T. Piran and I. Shemi, ApJ, {\bf 379},
L17 (1991); S.J. Curran, \& D.L. 
Lorimer, MNRAS {\bf 276}, 347 (1995); See, however,
L.V. Tutukov and A.R. Yungelston, MNRAS, {\bf
260}, 675 (1993); V.M. Lipunov et al., ApJ, {\bf 454}, 493 (1995).
 
21. See, e.g., E.P.J. van den Heuvel and D.R. Lorimer, MNRAS {\bf 283}, L37 
(1996); V.M. Lipunov et al., Ap. \& Sp. Phys. Rev. {\bf 9}, 1 (1996).

22. V.S. Berezinsky et al., {\it Ast. of Cosmic Rays},
(North Holland 1990) p. 71. 

23. C.C. Wu et al., ApJ. {\bf 416}, 247 (1993). 
 
24. A. Dar, A. Laor, \& N.J. Shaviv, to be published. 

25. J.W. Elbert, Proc. DUMAND Workshop (ed. A. Roberts) {\bf 2}, p. 101
(1978).

26. R.M. Barnett et al., Phys. Rev. D {\bf 54}, 1 (1996).
  
27. A. Dar, Phys. Rev. Lett. {\bf 51}, 227 (1983).

28. J.A.E. Stephenson et al., Nature {\bf 352}, 137 (1991).

29. C.B. Officer, \& C.L. Drake, Science, {\bf 227}, 1161 (1985). 

30. See, e.g., C.B. Officer et al., Nature {\bf 326}, 143 (1987).

31. See, e.g., L. Zhao \& F.T. Kyt, Earth planet Sci. Lett. {\bf 90}, 
411 (1988).

\end{document}